# Proximity Ferroelectricity in Compositionally Graded Structures

Eugene A. Eliseev[1*], Anna N. Morozovska[2*†], Sergei V. Kalinin[3], Long-Qing Chen[4‡], and Venkatraman Gopalan[4§]

[1]Frantsevich Institute for Problems in Materials Science, National Academy of Sciences of Ukraine,
3, str. Omeliana Pritsaka, 03142 Kyiv, Ukraine

[2] Institute of Physics, National Academy of Sciences of Ukraine,
46, Nauki avenue, 03028 Kyiv, Ukraine

[3]Department of Materials Science and Engineering, University of Tennessee,
Knoxville, TN, 37996, USA

[4] Department of Materials Science and Engineering,
Pennsylvania State University, University Park, PA 16802, USA

## Abstract

Proximity ferroelectricity is a novel paradigm for inducing ferroelectricity in a non-ferroelectric polar material such as AlN or ZnO that are typically unswitchable with an external field below their dielectric breakdown field. When placed in direct contact with a thin switchable ferroelectric layer (such as $Al_{1-x}Sc_xN$ or $Zn_{1-x}Mg_xO$), they become a practically switchable ferroelectric. Using the thermodynamic Landau-Ginzburg-Devonshire theory, in this work we performed the finite element modeling of the polarization switching in the compositionally graded $AlN$-$Al_{1-x}Sc_xN$, $ZnO$-$Zn_{1-x}Mg_xO$ and $MgO$-$Zn_{1-x}Mg_xO$ structures sandwiched in both a parallel-plate capacitor geometry as well as in a sharp probe-planar electrode geometry. We reveal that the compositionally graded structure allows the simultaneous switching of spontaneous polarization in the whole system by a coercive field significantly lower than the electric breakdown field of unswitchable polar materials. The physical mechanism is the depolarization electric field determined by the gradient of chemical composition "x". The field lowers the steepness of the switching barrier in the otherwise unswitchable parts of the compositionally graded $AlN$-$Al_{1-x}Sc_xN$ and $ZnO$-$Zn_{1-x}Mg_xO$ structures, while it induces a shallow double-well free energy potential in the MgO-like regions of compositionally graded $MgO$-$Zn_{1-x}Mg_xO$ structure. Proximity ferroelectric switching of the compositionally graded structures placed in the probe-electrode geometry occurs due to nanodomain formation under the tip. We predict that a gradient of chemical composition "x" significantly lowers effective coercive fields of the compositionally graded AlN-Al$_{1-x}$Sc$_x$N and ZnO-Zn$_{1-x}$Mg$_x$O structures compared to the coercive fields of the corresponding multilayers with a uniform chemical composition in each layer. A tip-induced switching further lowers the coercive field enabling control of ferroelectric domains in otherwise unswitchable compositionally graded structures can provide nanoscale domain control for memory, actuation, sensing and optical applications.

---

[*]These authors contributed equally

[†] corresponding author, e-mail: anna.n.morozovska@gmail.com

[‡] corresponding author, e-mail: lqc3@psu.edu

[§] corresponding author, e-mail: vgopalan@psu.edu

## I. Introduction

It is known that the atomic-scale chemical stress through bulk doping can lower dramatically the coercive field for ferroelectric switching in wurtzite to practical levels (as in Mg doping of polar but unswitchable ZnO, and Sc or B doping of the polar but unswitchable AlN) [1, 2, 3]. The recent observation of "proximity ferroelectricity" by Skidmore et al. [4] reveals experimentally the possibility of collective switching in wurtzite ferroelectric heterostructures. The layered structures, whose thicknesses varied from tens to hundreds of nm, included two-layer (asymmetric, e.g. Al$_{1-x}$Sc$_x$N/AlN, Al$_{1-x}$B$_x$N/AlN, ZnO/Al$_{1-x}$B$_x$N) and three-layer (symmetric, e.g. Al$_{1-x}$B$_x$N/AlN/Al$_{1-x}$B$_x$N, AlN/Al$_{1-x}$B$_x$N/AlN, Zn$_{1-x}$Mg$_x$O/ZnO/Zn$_{1-x}$Mg$_x$O) configurations [4].

The Landau-Ginzburg-Devonshire (LGD) theory of proximity ferroelectricity developed by Eliseev et. al. [5, 6] predicts regimes of both "proximity switching" where the multilayers collectively switch, as well as "proximity suppression" where they collectively do not switch. The mechanism of the proximity ferroelectricity is an internal electric field determined by the polarization of the layers and their relative thickness in a self-consistent manner that renormalizes the double-well ferroelectric potential to lower the steepness of the switching barrier. Further reduction in the coercive field emerges from charged defects in the bulk which act as nucleation centers, since correlated nucleation of the spike-like domains in the vicinity of sign-alternating randomly distributed electric charge sources [7], as well as correlated polarization switching in the proximity of ferroelectric domain walls [8], have a significant influence on the coercive field reduction.

Ban et al. [9] developed a methodology for analyzing compositionally graded ferroic materials based on the generalized LGD approach. Material system inhomogeneities are assumed to arise from compositional, temperature, strain or stress gradients. Compositionally graded ferroelectrics produce internal potentials that manifest themselves as asymmetric hysteresis in polarization versus applied field plots [10]. Compositionally graded ferroelectrics, as wide band gap semiconductors, reveal electrical domain structures [11]. Silicon-compatible wurtzite compositionally graded ferroelectrics are very promising for advanced nanoelectronics, optoelectronics and related emerging technologies.

To the best of our knowledge, proximity ferroelectricity has not been studied in the compositionally graded wurtzite ferroelectric nanostructures. Using the LGD thermodynamical theory, in this work we performed the finite element modeling (FEM) of the polarization switching in the compositionally graded AlN-Al$_{1-x}$Sc$_x$N, ZnO-Zn$_{1-x}$Mg$_x$O and MgO-Zn$_{1-x}$Mg$_x$O structures sandwiched in the parallel-plate capacitor and in the probe-planar electrode geometry. We reveal that the proximity of the "otherwise unswitchable" polar materials (such as AlN and ZnO) or dielectrics (like MgO) to the switchable ferroelectrics (such as Al$_{1-x}$Sc$_x$N and Zn$_{1-x}$Mg$_x$O) in the compositionally graded structure allows the simultaneous switching of spontaneous polarization in the whole structure. We obtained that a chemical composition gradient lowers the effective coercive fields of the compositionally graded AlN-Al$_{1-x}$Sc$_x$N and ZnO-Zn$_{1-x}$Mg$_x$O structures compared to the coercive fields of the defect-free AlN-Al$_{1-x}$Sc$_x$N and ZnO-Zn$_{1-x}$Mg$_x$O multilayers with a uniform chemical composition in each layer, where the proximity ferroelectricity was discovered earlier [4, 5, 6]. The compositionally graded MgO-Zn$_{1-x}$Mg$_x$O structure sandwiched in a parallel-plate capacitor has a much smaller effective coercive field and a narrow antiferroelectric (AFE) type polarization hysteresis, indicating low power consumption for complete switching. The probe-planar electrode geometry lowers the coercive field of the graded structure even more, though a small hysteresis appears in this case.

## II. Problem Statement

We assume that the chemical composition has a pronounced gradient near the electrically open surface of wurtzite ferroelectrics, such as Al$_{1-x}$Sc$_x$N or Zn$_{1-x}$Mg$_x$O. The gradient-type changes of the chemical composition can emerge due to the adsorption of ions and/or vacancies during electrochemical reactions, which can take place under the biased tip of the PFM probe [12, 13]. The artificial gradient of the chemical composition can be created too. We consider the situation, when the compositionally graded layer is an irreversible polar piezoelectric (AlN or ZnO), or a paraelectric or dielectric (MgO), which gradually transforms into the uniaxial ferroelectric Al$_{1-x}$Sc$_x$N or Zn$_{1-x}$Mg$_x$O, respectively. The total thickness of the compositionally graded structure is $h$. The thickness of the chemical composition gradient layer is determined by several diffusion lengths $h_d$ in the exponential decay factor.

A sketch of the compositionally graded structure sandwiched between the PFM tip and the planar electrode is shown in **Fig. 1(a)**. The tip-surface contact is considered using the model of "shielded" disk-probe that allows to fix the electric potential $\varphi$ at the flat surface $z = h$ (see e.g., Refs. [13, 14]). Within the model, the probe apex is approximated by a biased disk of the radius $R \sim (2.5 - 5)$ nm. The free surface of the wurtzite layer outside the disk electrode is covered by a layer of screening charge (not shown in the figure), which is capable to provide the condition $\varphi = 0$ outside the probe-surface contact.

For comparison, we consider the case of the top planar electrode ($R \to \infty$). A sketch of the compositionally graded structure sandwiched between the parallel-plate electrodes is shown in **Fig. 1(b)**.

A periodic electric voltage $U(t)$ applied to the tip (or to the top planar electrode), has an amplitude increasing linearly in time from zero to $U_{max}$, namely $U(t) = U_{max} \frac{t}{t_{max}} sin(\omega t)$, where $\omega$ is the pulse frequency and $t_{max}$ is the pulse duration. The substrate electrode is planar and regarded as electrically grounded ($\varphi = 0$ at $z = 0$).

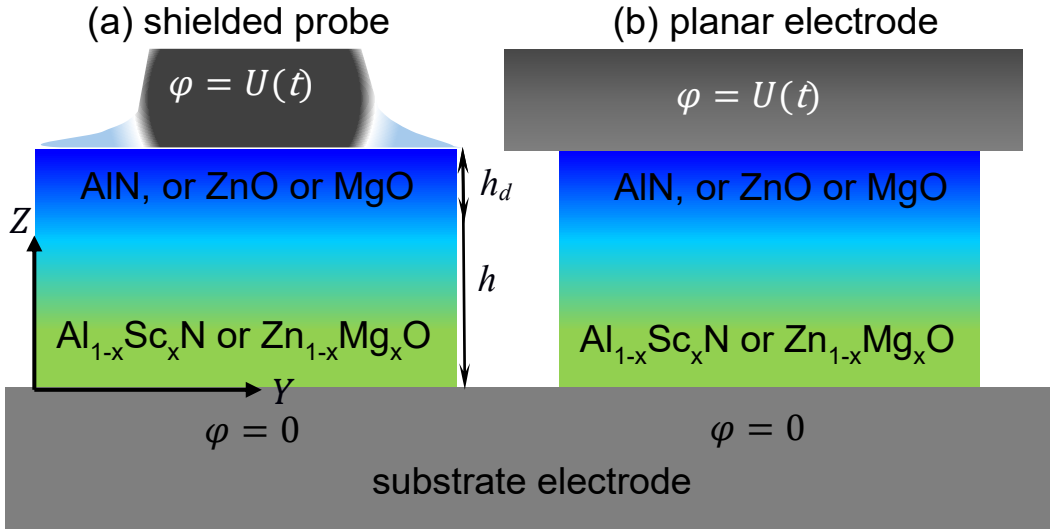

**FIGURE 1**. Considered heterostructures: the "probe – compositionally graded ferroelectric (AlN-Al$_{1-x}$Sc$_x$N, or ZnO-Zn$_{1-x}$Mg$_x$O or MgO-Zn$_{1-x}$Mg$_x$O) – substrate electrode" **(a)**, and the "top planar electrode – compositionally graded ferroelectric (AlN-Al$_{1-x}$Sc$_x$N, or ZnO-Zn$_{1-x}$Mg$_x$O or MgO-Zn$_{1-x}$Mg$_x$O) – substrate electrode" **(b)**. The total thickness of the ferroelectric structure is $h$ and the thickness of the compositionally graded polar (or dielectric) layer is about $3h_d$. The sample coordinates {X, Y, Z} form a right-handed coordinate system.

Importantly, that the shielded probe (or the top planar electrode) is regarded as "elastically soft" in order not to perturb the local piezoelectric response of the wurtzite film. The substrate electrode is regarded "elastically rigid". The elastic and electric boundary conditions are periodic on the side boundaries of the computational cell.

The time-dependent LGD equation for the ferroelectric polarization $P_z$ inside compositionally graded structure is the following:

$$\Gamma \frac{\partial P_z}{\partial t} + \alpha(z)P_z + \beta(z)P_z^3 + \gamma(z)P_z^5 - g_z \frac{\partial^2 P_z}{\partial z^2} - g_\perp \left( \frac{\partial^2 P_z}{\partial x^2} + \frac{\partial^2 P_z}{\partial y^2} \right) = E_z(x, y, z). \quad (1a)$$

Here $\Gamma$ is the Landau-Khalatnikov relaxation coefficient. $E_z$ is z-component of electric field acting inside the structure. The Landau expansion coefficients $\alpha$, $\beta$ and $\gamma$ depend on $z$ due to the z-gradient of the chemical composition "$x$" in the following way:

$$\alpha(z) = \alpha_1 + (\alpha_2 - \alpha_1)\exp\left(-\frac{h-z}{h_d}\right) - 2Q_{13}(z)(\sigma_{22} + \sigma_{11}) - 2Q_{33}(z)\sigma_{33}, \quad (1b)$$

$$\beta(z) = \beta_1 + (\beta_2 - \beta_1)\exp\left(-\frac{h-z}{h_d}\right), \quad \gamma(z) = \gamma_1 + (\gamma_2 - \gamma_1)\exp\left(-\frac{h-z}{h_d}\right). \quad (1c)$$

Here the coefficients $\alpha_1$, $\beta_1$ and $\gamma_1$ correspond to the ferroelectric (Al$_{1-x}$Sc$_x$N or Zn$_{1-x}$Mg$_x$O); the coefficients $\alpha_2$, $\beta_2$ and $\gamma_2$ correspond to the piezoelectric (AlN or ZnO) or dielectric MgO. The coefficient $\alpha(z)$ is renormalized by elastic stresses $\sigma_{ij}$ via the electrostriction coefficients $Q_{ij}$. For the sake of simplicity, we regard that elastic constants are the same in the gradient-type polar (or dielectric) layer and in the pristine ferroelectric layer. Elastic stresses satisfy the equation of mechanical equilibrium in the computation region, $\frac{\partial \sigma_{ij}}{\partial x_j} = 0$. Elastic equations of state follow from the variation of the free energy with respect to elastic stress, namely:

$$s_{ijkl}\sigma_{ij} + Q_{ijkl}P_k P_l = u_{ij}. \quad (1c)$$

Elastic boundary conditions correspond to the absence of normal stresses at the top surface $z = h$, and zero elastic displacement at the rigid substrate $z = 0$. Hereafter, we neglect the influence of the flexoelectric coupling for the sake of simplicity.

For this work, we regard the normal derivative of the polarization to be a continuous function inside the graded structure and to be zero at the top and bottom surfaces:

$$\left.\frac{\partial P_z^{(1)}}{\partial z}\right|_{z=0} = 0, \qquad \left.\frac{\partial P_z^{(2)}}{\partial z}\right|_{z=h} = 0. \quad (2)$$

The electric field $E_z = -\frac{\partial \varphi}{\partial z}$, where the electric potential $\varphi$ obeys the Poisson equation:

$$\varepsilon_0 \varepsilon_b \left(\frac{\partial^2}{\partial x^2} + \frac{\partial^2}{\partial y^2} + \frac{\partial^2}{\partial z^2}\right)\varphi = \frac{\partial^2 P_z}{\partial z^2}. \quad (3)$$

Here $\varepsilon_0$ is a universal dielectric constant, and $\varepsilon_b(z) = \varepsilon_b^{(1)} + \left(\varepsilon_b^{(2)} - \varepsilon_b^{(1)}\right)\exp\left(-\frac{h-z}{h_d}\right)$, where $\varepsilon_b^{(1)}$ and $\varepsilon_b^{(2)}$ correspond to the background permittivity [15] of ferroelectric and piezoelectric (or dielectric) materials, respectively. The electric boundary conditions are the fixed potential at the electrodes,

$$\varphi = 0|_{z=0}, \qquad \varphi = U(x,y,t)|_{z=h}. \quad (4)$$

LGD parameters of Al$_{1-x}$Sc$_x$N and Zn$_{1-x}$Mg$_x$O, used in the FEM, are listed in **Table SI** in **Supplementary Materials**. The LGD parameters and dielectric constants of Al$_{1-x}$Sc$_x$N were determined from Refs. [16, 17, 18, 19]. The LGD parameters and dielectric constants of Zn$_{1-x}$Mg$_x$O were determined from Refs. [20, 21, 22, 23]. We assume equal gradient tensor coefficients in longitudinal and transverse directions. Electrostriction coefficients $Q_{ij}$ and elastic stiffness tensors $c_{ij}$ of the ferroelectric, polar and/or dielectric materials are collected from Refs. [24, 25]. They are listed in **Tables S2-S3** in **Supplementary Materials.** The procedure for the LGD parameters determination is described in detail in Refs. [5, 6].

### III. FEM Results and Discussion

We performed the FEM of the polarization, electric and elastic fields evolution in the compositionally graded AlN-Al$_{0.73}$Sc$_{0.27}$N, ZnO-Zn$_{0.63}$Mg$_{0.37}$O and MgO-Zn$_{0.63}$Mg$_{0.37}$O structures sandwiched in the parallel-plate capacitor and in the probe-planar electrode geometry, shown in **Fig. 1(b)** and **1(a)**, respectively. We use a uniform rectangular mesh with the size 2 nm, and the linear size of the cubic computation cell is 40 nm. The diffusion length $h_d$ varies from 3 to 10 nm. The choice of $h_d$ range corresponds to the situation, when a significant part of the graded structure with thickness about several $h_d$ has LGD coefficients close to the otherwise unswitchable polar materials AlN or ZnO, or dielectric MgO. Due to the diffusion length, the chemical composition "x" continuously changes from 0 to 0.27 for the compositionally graded AlN-Al$_{0.73}$Sc$_{0.27}$N structure; or from 0 to 0.37 for the compositionally graded ZnO-Zn$_{0.63}$Mg$_{0.37}$O and MgO-Zn$_{0.63}$Mg$_{0.37}$O structures. The x-changes lead to the gradual changes of the LGD coefficients $\alpha(z)$, $\beta(z)$ and $\gamma(z)$ in accordance with Eqs.(1b) and (1c).

The initial distribution of polarization is a single-domain state with randomly small fluctuations for the compositionally graded AlN-Al$_{0.73}$Sc$_{0.27}$N, ZnO-Zn$_{0.63}$Mg$_{0.37}$O structures; and the paraelectric state for the compositionally graded MgO-Zn$_{0.63}$Mg$_{0.37}$O structure. Initial states relax to equilibrium distributions, and, after the complete relaxation, a periodic sinusoidal bias voltage $U(t)$, whose amplitude increases linearly in time, is switched on between the electrodes. The period of applied voltage is four orders of magnitude higher than the Landau-Khalatnikov relaxation time of polarization.

### A. Proximity ferroelectricity in the compositionally graded ferroelectrics sandwiched in the parallel-plate capacitor

First, let us analyze the FEM results for the compositionally graded AlN-Al$_{0.73}$Sc$_{0.27}$N and ZnO-Zn$_{0.63}$Mg$_{0.37}$O structures sandwiched between the parallel-plate electrodes (as shown in **Fig. 1(b)**). Since the graded top layer, whose chemical composition is close to the AlN (or ZnO) at $h < z < h - h_d$, supports the polarized state, the ground state of the compositionally graded AlN-Al$_{0.73}$Sc$_{0.27}$N and ZnO-Zn$_{0.63}$Mg$_{0.37}$O structures are single-domain at zero bias. Equilibrium z-profiles of the spontaneous polarization $P_z$ inside the compositionally graded structures are shown by solid curves in **Figs. 2(a)**, respectively. Dashed horizontal lines show the average polarization $\bar{P}_z$. As follows from the small difference between $P_z(z)$ and $\bar{P}_z$, the single-domain spontaneous polarization is almost homogeneous inside the structures. Z-profiles of the depolarization field $E_z$ in the structures are shown by the blue and green curves in **Fig. 2(b)**. The depolarization field $E_z$, whose direction is opposite to $P_z$ in the part of the graded structure with the larger spontaneous polarization (e.g., where the chemical composition "x" is close to AlN) and coincides with $P_z$ in the part of the structure with the lower spontaneous polarization (e.g., where the chemical composition "x" is close to $Al_{0.73}Sc_{0.27}N$), equalizes the spontaneous polarization in entire the graded structure. The resulting distribution of the spontaneous polarization is almost homogeneous inside the structure disregarding the chemical composition gradient. In other words, the depolarization field almost suppresses the influence of the chemical composition gradient on the spontaneous polarization. This behavior of the polarization is in a qualitative agreement with its behavior in the wurtzite bilayers, considered in Refs. [5], where the depolarization field equates the polarization of the AlN (or ZnO) and $Al_{0.73}Sc_{0.27}N$ (or $Zn_{0.63}Mg_{0.37}O$) layers to minimize the electrostatic energy of the system. The same mechanism equates the spontaneous polarization in the three-layers [6] and, by extension, in multilayers. As a matter of fact, compositionally graded structures, considered in this work, can be represented as a continuous set of atomically thin layers, whose chemical composition "x" changes on a small value Δx in each layer. This representation helps to explain the mechanism of the depolarization field influence.

The amplitude of the sinusoidal bias voltage increases linearly in time as shown in **Fig. 2(c)**. Time sweeps of the average polarization $\bar{P}_z$ in the compositionally graded structures are strongly anharmonic periodic functions with pronounced vertical lines corresponding to the single-domain polarization switching, which happens when the voltage amplitude exceeds relatively high coercive values (see **Figs. 2(d)** and **2(e)**). The sweeps correspond to the rectangular hysteresis loops of $\bar{P}_z$ shown in **Figs. 2(f)** and **2(g)**, respectively. The coercive voltages 43 V (for AlN-$Al_{0.73}Sc_{0.27}N$) and 40 V (for ZnO-$Zn_{0.63}Mg_{0.37}O$) correspond to the coercive fields 10.8 MV/cm and 10 MV/cm, respectively. These fields are significantly lower than the hypothetic thermodynamic coercive fields of "otherwise unswitchable" AlN and ZnO, which are equal to 26.1 MV/cm and 24.8 MV/cm, respectively, being much higher than the electric breakdown fields of the materials (see **Table SI**). At the same time the coercive fields of the compositionally graded AlN-$Al_{0.73}Sc_{0.27}N$ and ZnO-$Zn_{0.63}Mg_{0.37}O$ structures are a bit higher than the thermodynamic coercive fields of the bulk $Al_{0.73}Sc_{0.27}N$ (9.3 MV/cm) and $Zn_{0.63}Mg_{0.37}O$ (8.9 MV/cm) materials, respectively.

On the other hand, the coercive fields of the compositionally graded AlN-$Al_{0.73}Sc_{0.27}N$ and ZnO-$Zn_{0.63}Mg_{0.37}O$ structures, which are 10.6 MV/cm and 10 MV/cm, are significantly smaller than the coercive fields of defect-free AlN-$Al_{0.73}Sc_{0.27}N$ and ZnO-$Zn_{0.63}Mg_{0.37}O$ bilayers with a uniform chemical composition of each layer, which are 17.7 MV/cm and 14.7 MV/cm, respectively (see **Table S4**, Fig. 4(c) and Fig. 7(c) in Ref. [5]). Note that the comparison of coercive fields is relevant only when the ratio of the effective thickness of the chemical gradient (~$3h_d$) to the total thickness $h$ of the compositionally graded structure is the same as the ratio of polar (AlN or ZnO) layer thickness $h_2$ to the total thickness $h$ of the bilayer. A possible reason for the significant reduction in the coercive field of the compositionally graded structure compared to the corresponding bilayer is the physical analogy between the gradient of chemical composition "x", which induces a gradient of chemical strains, and defect-induced fields, since defects are known to reduce significantly the coercive field (compare e.g., the coercive field in Fig. 4(c) and Fig. 5(b) in Ref. [5]).

We also note that the average spontaneous polarization $\bar{P}_S$ of the compositionally graded AlN-Al$_{0.73}$Sc$_{0.27}$N and ZnO-Zn$_{0.63}$Mg$_{0.37}$O structures (110 μC/cm$^2$ and 90 μC/cm$^2$) is slightly smaller than $\bar{P}_S$ of defect-free AlN-Al$_{0.73}$Sc$_{0.27}$N and ZnO-Zn$_{0.63}$Mg$_{0.37}$O bilayers (120 μC/cm$^2$ and 96 μC/cm$^2$) (see **Table S4**). The reason for the decrease in $\bar{P}_S$ is the influence of chemical strains on the position of the free energy potential double-wells, which determines the value of $\bar{P}_S$. The influence of the chemical strains on $\bar{P}_S$ is smaller in comparison with its stronger influence on the barrier height, which mainly determines the value of thermodynamic coercive field. Thus, the gradient of chemical composition allows us to create compositionally graded wurtzite structures with a significantly lower coercive field and a slightly smaller spontaneous polarization compared to compositionally uniform multilayers.

Thus, the proximity of the unswitchable polar materials (such as AlN and ZnO) to the switchable ferroelectrics (such as Al$_{1-x}$Sc$_x$N and Zn$_{1-x}$Mg$_x$O) in the compositionally graded structure allows the simultaneous switching of the spontaneous polarization in the entire structure by a coercive field significantly lower than the electric breakdown field in the unswitchable polar materials. The physical mechanism of the proximity ferroelectricity [4] in the considered compositionally graded AlN-Al$_{1-x}$Sc$_x$N and ZnO-Zn$_{1-x}$Mg$_x$O structures is a depolarization electric field determined by the gradient of $P_z$ in the normal direction z (see e.g., **Fig. 2(b)**). The field continuously renormalizes the double-well potential of the free energy to lower the steepness of the switching barrier in the "otherwise unswitchable" parts of the graded structure. In result, the depolarization field, that counteracts the gradient of chemical composition "x", supports the single-domain polarization switching in the whole graded structure. Remarkably, that the x-gradient significantly lowers effective coercive fields of the compositionally graded structures compared to the coercive fields of the multilayers with a uniform chemical composition of each layer. At the same time the influence of the composition gradient on the average spontaneous polarization of the structures is smaller.

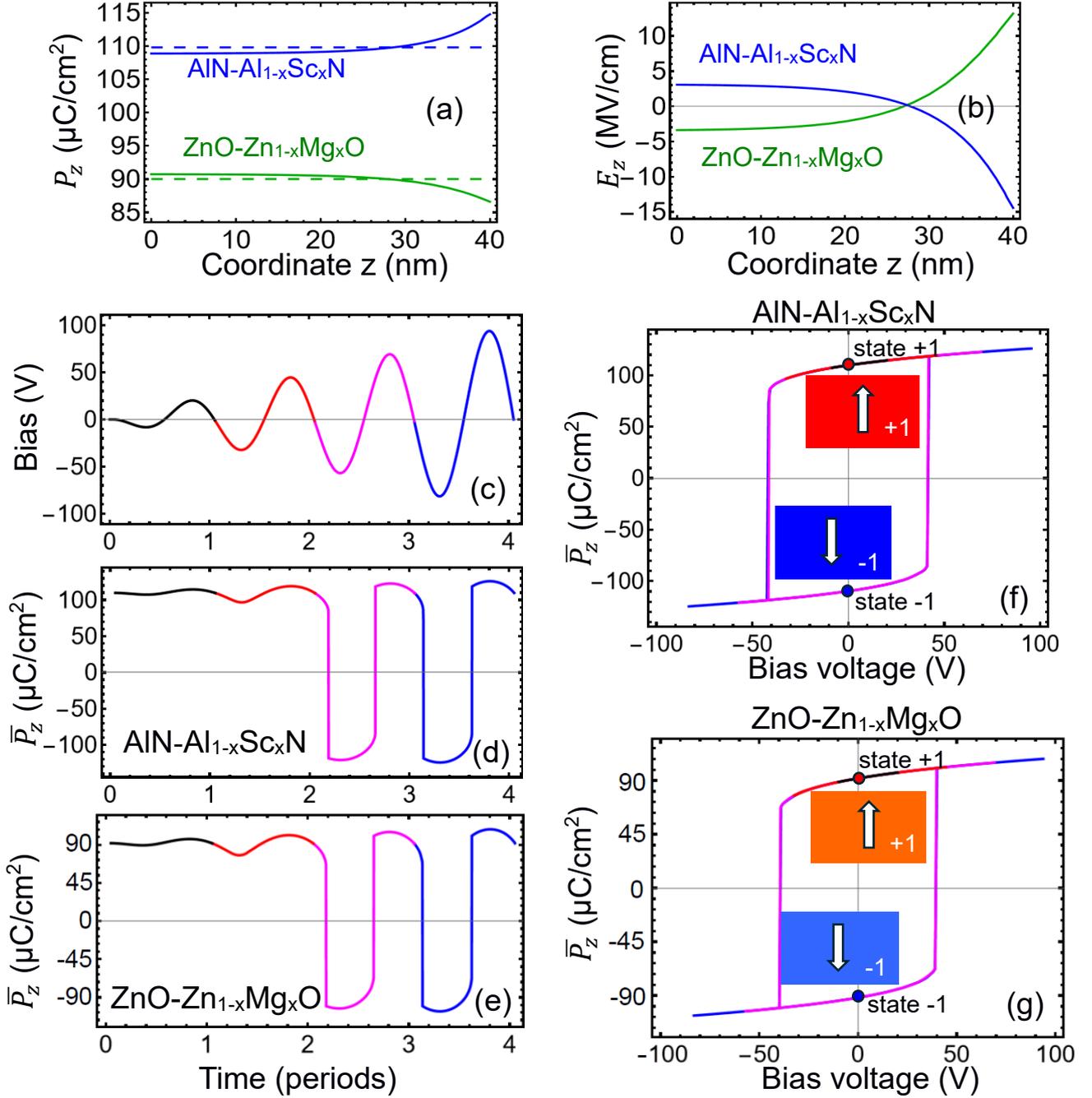

**FIGURE 2**. Z-profiles of the spontaneous polarization $P_z$ (a) and the depolarization field $E_z$ (b) in the ground state of the compositionally graded AlN-Al$_{0.73}$Sc$_{0.27}$N (blue solid curves) and ZnO-Zn$_{0.63}$Mg$_{0.37}$O (green solid curves) structures sandwiched between the short-circuited parallel-plate electrodes ($U = 0$). Dashed lines in the plot (a) show the average spontaneous polarization. Time dependence of the bias voltage applied between the electrodes (c). The time (d, e) and bias voltage (f, g) dependences of the average polarization $\bar{P}_z$ in the compositionally graded AlN-Al$_{0.73}$Sc$_{0.27}$N (d, f) and ZnO-Zn$_{0.63}$Mg$_{0.37}$O (e, g) structures. The structure thickness $h = 40$ nm and the diffusion length $h_d = 7$ nm. Material parameters are listed in **Tables S1-S3**.

Approximate analytical expressions for the average spontaneous polarization and effective coercive field can be calculated in the following way. Using the boundary conditions (2), the average spontaneous polarization $\bar{P}_S$ can be estimated after the averaging of Eq.(1a) over z-coordinate, which yields the equation

$$\bar{\alpha}\bar{P}_S + \bar{\beta}\bar{P}_S^3 + \bar{\gamma}\bar{P}_S^5 = 0, \quad (5a)$$

where we use the approximation $\overline{\alpha P_S} \approx \bar{\alpha} \cdot \bar{P}_S$, $\overline{\beta P_S^3} \approx \bar{\beta} \cdot \bar{P}_S^3$ and $\overline{\gamma P_S^5} \approx \bar{\gamma} \cdot \bar{P}_S^5$; and introduce the average coefficients $\bar{\alpha}$, $\bar{\beta}$ and $\bar{\gamma}$:

$$\bar{\alpha} = \frac{1}{h}\int_0^h \alpha(z)\, dz = \alpha_1 + (\alpha_2 - \alpha_1)\frac{h_d}{h}\left[1 - \exp\left(-\frac{h}{h_d}\right)\right], \quad (5b)$$

$$\bar{\beta} = \frac{1}{h}\int_0^h \beta(z)\, dz = \beta_1 + (\beta_2 - \beta_1)\frac{h_d}{h}\left[1 - \exp\left(-\frac{h}{h_d}\right)\right], \quad (5c)$$

$$\bar{\gamma} = \frac{1}{h}\int_0^h \gamma(z)\, dz = \gamma_1 + (\gamma_2 - \gamma_1)\frac{h_d}{h}\left[1 - \exp\left(-\frac{h}{h_d}\right)\right]. \quad (5d)$$

The average spontaneous polarization is given by expression:

$$\bar{P}_S \approx \frac{1}{2\bar{\gamma}}\left(-\bar{\beta} + \sqrt{\bar{\beta}^2 - 4\bar{\alpha}\bar{\gamma}}\right). \quad (6a)$$

The thermodynamic coercive field $E_c^{th}$ can be estimated assuming that the homogeneous electric field $U/h$ stands in the right-hand part of Eq.(5a). Thus

$$E_c^{th} = \frac{2}{5}\left(2\bar{\beta} + \sqrt{(3\bar{\beta})^2 - 20\bar{\alpha}\bar{\gamma}}\right)\left(\frac{2\bar{\alpha}}{-3\bar{\beta} - \sqrt{(3\bar{\beta})^2 - 20\bar{\alpha}\bar{\gamma}}}\right)^{3/2}. \quad (6b)$$

Expressions (6) with the average coefficients $\bar{\alpha}$, $\bar{\beta}$ and $\bar{\gamma}$ describe quantitatively the behavior of spontaneous polarization and coercive field in the compositionally graded AlN-Al$_{1-x}$Sc$_x$N and ZnO-Zn$_{1-x}$Mg$_x$O structures shown in **Fig. 2(f)** and **2(g)**.

Next, we analyze the FEM results for the compositionally graded MgO-Zn$_{0.63}$Mg$_{0.37}$O structure sandwiched between the parallel-plate electrodes (shown in **Fig. 3**). Since the top layer, whose chemical composition is close to the MgO at $h < z < h - h_d$, acts as a diffuse dielectric gap, which creates a strong depolarization field, the compositionally graded structure splits into the vertical domain stripes at zero bias voltage (see **Fig. 3(a)**). It is important that the low-contrast domain stripes exist in the top layer $h < z < h - 3h_d$, being induced by the proximity of the dielectric MgO to the ferroelectric Zn$_{0.63}$Mg$_{0.37}$O. However, the stripes of $P_z$ lose their contrast gradually and vanish approaching the MgO top surface $z = h$. A very specific distribution of the depolarization field $E_z$ determines the stability of the up-polarized and down-polarized domain stripes near the MgO surface of the graded structure.

Indeed, the distribution of $E_z$ (shown in **Fig. 3(b)**) has the view of high-contrast vertical stripes with opposite polarity in the top layer $h < z < h - 2h_d$, reaches minimum at $z \approx h - 3h_d$, changes the polarity and gradually loses the contrast with decrease in z coordinate, and then appears again in the form of small circles with counter-polarity near the bottom electrode $z \approx 0$. Thus, the depolarization field is concentrated in the graded layer near the top electrode, where it induces and supports the low-contrast domain stripes in MgO. The field behavior near the electrodes is influenced by the image charges.

Using a direct variational method for the determination of the equilibrium period $\frac{2\pi}{k_{eq}}$ of the domain stripes, as proposed in Ref. [26], we obtained the approximate analytical expression for $k_{eq}$:

$$k_{eq} \cong \begin{cases} 0, & 0 \leq h_d < h_{cr}, \\ \sqrt{\frac{2}{h\, h_{cr}} \frac{\varepsilon_b^{(2)}}{\varepsilon_b^{(1)}} - \frac{2}{h^2}\left(1 + \frac{\varepsilon_b^{(2)} h}{\varepsilon_b^{(1)} h_d}\right)}, & h_{cr} \leq h_d \leq 10 h_{cr}, \\ \sqrt{\frac{\pi}{\sqrt{2} h h_{cr}} \frac{\varepsilon_b^{(2)}}{\varepsilon_b^{(1)}} - \frac{\pi^2}{4h^2} - 2 \frac{\varepsilon_e}{\varepsilon_b^{(1)} h\, h_d}}, & h_d > 10 h_{cr}. \end{cases} \quad (7)$$

Here $h_{cr} \approx 2 \frac{\varepsilon_b^{(2)}}{\varepsilon_b^{(1)}} \sqrt{2\varepsilon_0 \varepsilon_b^{(1)} g_\perp}$ is the critical diffusion length above which the domain splitting starts at zero bias voltage. The condition $h_d < h_{cr}$ is required for the absolute stability of a single-domain state of the whole structure. The condition $h_d \gtrsim h_{cr}$ corresponds to the appearance of the very broad domains with a small period. The "thick" dielectric layer $h_d \gg h_{cr}$ induces domain stripes with the width proportional to $\sqrt{h}$ (see e.g., Ref. [26]). Expressions (7) quantitatively describe the dependence of the domain stripes period on the thickness $h$ and $h_d$, as well as the value of $h_{cr}$. In particular, $h_{cr}$ is less than 1 nm in the considered compositionally graded MgO-Zn$_{0.63}$Mg$_{0.37}$O structure, being comparable to a physical dielectric gap.

The time sweep of the average polarization $\bar{P}_z$ in the compositionally graded MgO-Zn$_{0.63}$Mg$_{0.37}$O structure is a slightly anharmonic periodic function corresponding to the polydomain polarization switching (see **Figs. 3(d)**). The sweep corresponds to the very thin AFE-like hysteresis loops of $\bar{P}_z$ shown in **Fig. 3(e)**. Such type of poly-domain polarization switching leads to the virtual absence of coercivity, at the same time minimizing energy dissipation, while being able to switch. For the AFE-like loop shown in **Fig. 3(e)**, the dissipation energy density is 19 MPa at the field 24 V/nm (the area of the double-loop), while the stored energy is 164 MPa (the area above the double-loop).

Thus, the proximity ferroelectricity in the compositionally graded MgO-Zn$_{0.63}$Mg$_{0.37}$O structure is confirmed by the presence of the counter-polarized domain stripes in its top part, whose chemical composition is close to the dielectric MgO. Indeed, the distributions of both spontaneous polarization and depolarization field, prove the stability of the up-polarized and down-polarized domain stripes in the structure. However, a rectangular-shaped hysteresis loop with a pronounced coercive voltage is absent. The loop can appear under conditions $h \gg h_d$ or $h_d < h_{cr}$, where $h_{cr} < 1$ nm.

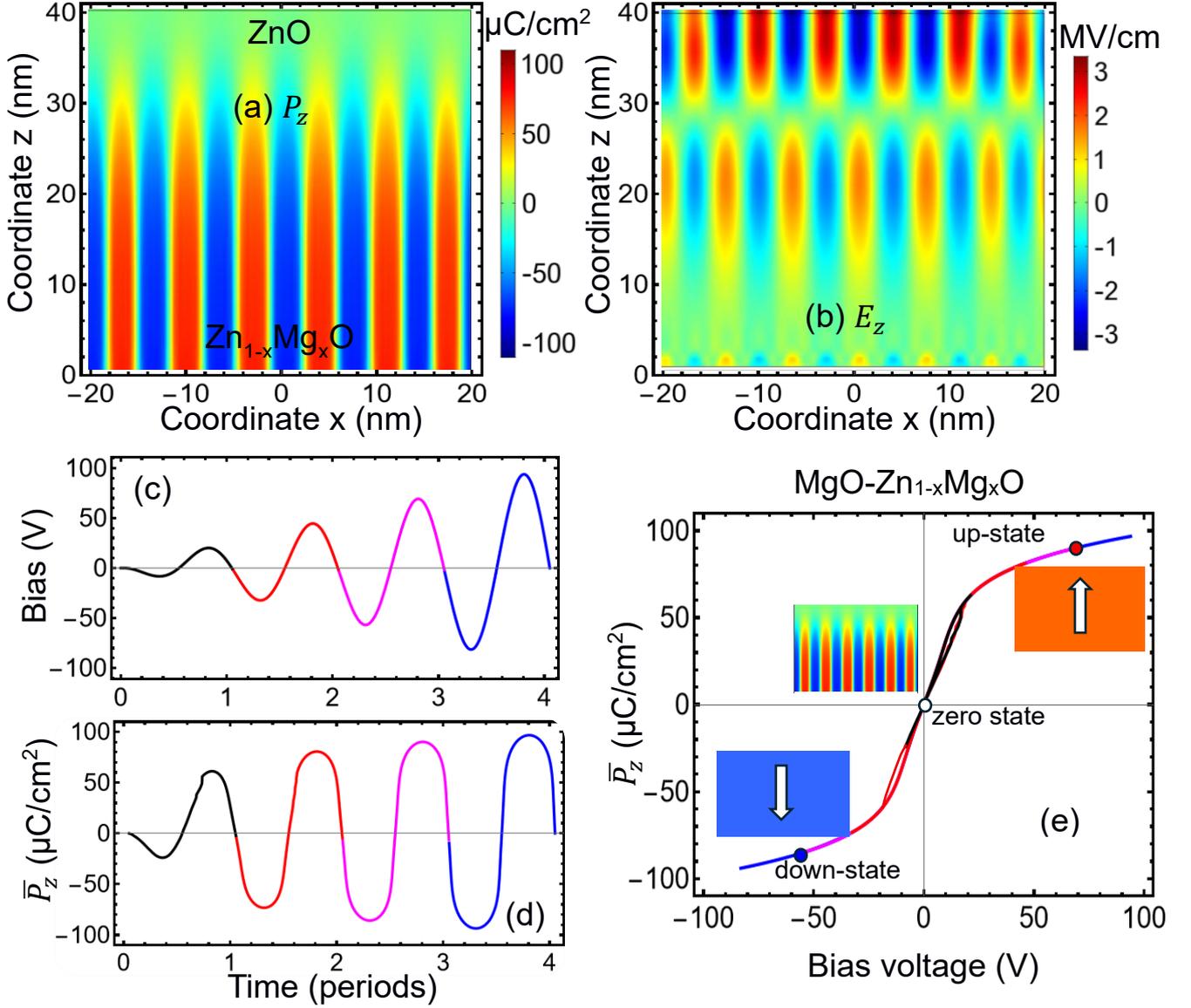

**FIGURE 3**. Distributions of the spontaneous polarization $P_z$ **(a)** and the depolarization field $E_z$ **(b)** in the ground state of the compositionally graded MgO-Zn$_{0.63}$Mg$_{0.37}$O structure sandwiched between the short-circuited parallel-plate electrodes ($U = 0$). Time dependence of the bias voltage applied between the electrodes **(c)**. The time **(d)** and bias voltage **(e)** dependences of the average polarization $\bar{P}_z$ in the graded structure. The structure thickness $h = 40$ nm and the diffusion length $h_d = 3$ nm. Material parameters are listed in **Tables S1-S3**.

The physical mechanism of the proximity ferroelectric switching [4] in the compositionally graded MgO-Zn$_{1-x}$Mg$_x$O structure is the depolarization electric field, which is determined by the polarization gradient components $\frac{\partial P_z(x,z)}{\partial z}$ and $\frac{\partial P_x(x,z)}{\partial x}$ (see e.g., **Fig. 3(b)**, where is $E_z(x,z)$ shown). The field continuously renormalizes the parabolic potential of the free energy in the top layer $h < z < h - 3h_d$ towards the appearance of the shallow double-well potential in the "otherwise dielectric" parts of the graded structure. Thus, the depolarization field, reinforced by the gradient of chemical composition "x", induces the formation of thermodynamically stable of low-contrast or high-contrast domain stripes in the top and bottom parts of the graded structure, respectively.

### B. Proximity-induced ferroelectric switching in the compositionally graded structures in the probe-plate electrode geometry

Below we analyze the polarization switching in the compositionally graded AlN-Al$_{0.73}$Sc$_{0.27}$N and ZnO-Zn$_{0.63}$Mg$_{0.37}$O structures placed between the PFM probe and the bottom planar electrode (the geometry is shown in **Fig. 1(a)**). Due to the very inhomogeneous distribution of the electric field in the probe-planar electrode geometry, where the field is concentrated in a small region under the PFM probe for the tip radius $R < 5$ nm, the local polarization and piezoelectric response are sensitive to the field changes under the probe only. The polarization distribution is almost homogeneous far from the probe, since the completely screened top surface is equivalent to the planar electrode (see **Figures 4(g) and 5(g)**). The homogeneous polarization distribution far from the probe coincides with the polarization distribution in the parallel-plate capacitor geometry (shown in **Fig. 2**). It should be noted that if we swap the AlN (or ZnO) and Al$_{0.73}$Sc$_{0.27}$N (or Zn$_{0.63}$Mg$_{0.37}$O) layers, the coercive voltage increases, similarly to what has been shown previously for bilayers [6]. The physical reason for the dependence of local polarization reversal and piezoelectric response on the layer-swapping is the opposite direction of the depolarization field in the graded part and in the rest of the structure, as well as its strong inhomogeneity under the biased probe. The depolarization field renormalizes the double-well ferroelectric potential to lower the steepness of the switching barrier in the otherwise "unswitchable" AlN and ZnO graded parts. Note that this effect is intrinsic to both compositionally graded structures and multilayers, being independent of any additional reduction in coercive field due to the presence of defects [6].

The amplitude of bias voltage increases in time linearly as shown in **Figs. 4(a) and 5(a)**. The time sweeps of the vertical surface displacement $u_z$ under the PFM tip are shown in **Fig. 4(b) and 5(b)**. The time sweeps of the polarization $z$-component $\overline{P}_z$ averaged over the volume of the computation cell, namely in the cylinder $\{0 \leq \sqrt{x^2 + y^2} < 4R,\ 0 \leq z \leq h\}$, are shown in **Fig. 4(c) and 5(c).** The time sweeps of $u_z$ and $\overline{P}_z$ are strongly anharmonic periodic functions with pronounced vertical sections corresponding to the polarization switching under the probe (similar to the polarization switching in the parallel-plate capacitor geometry shown in **Fig. 2**). The bias voltages below the coercive voltage are insufficient to induce the polarization reversal under the probe (see black curves in **Fig. 4(c) and 5(c)**). When the voltage overcomes the coercive value, the polarization reversal occurs under the PFM probe (see red, magenta and blue curves in **Fig. 4(c) and 5(c)**) and corresponding hysteresis loops of local polarization and piezoresponse appear, as shown **Fig. 4(d)-(f) and 5(d)-(f)**.

Notably that the up shift of $\bar{P}_z$ sweeps, shown in **Figs. 4(c)** and **5(c)**, is conditioned by the polarization averaging over the whole computation cell, which underestimates significantly the complete switching of polarization under the probe tip (as shown by blue domains in **Figs. 4(g)** and **5(g)**). A vertical shift is absent for the parallel-plate capacitor geometry (compare the sweeps in **Fig. 2(d)** and **2(e)** with those in **Figs. 4(g)** and **5(g)**).

Since the local curvature of the surface induced by the piezoelectric effect contains the response from the areas far from the probe, the butterfly-like loops of the surface displacement $u_z$, shown in **Figs. 4(d)** and **5(d)**, are asymmetric. In contrast, the local response to the changes of the probe electric field is symmetric. Namely, the hysteresis loops of the local piezoresponse phase are symmetric, as shown in **Figs. 4(e)** and **5(e)**. The hysteresis loops of the local polarization $\bar{P}_z^{local}$ averaged over a small region under the tip, namely in the cylinder $\{0 \leq \sqrt{x^2 + y^2} < R,\ 0 \leq z \leq h\}$, are symmetric too, as shown in **Figs. 4(f)** and **5(f)**.

The coercive voltages, corresponding to the probe-planar electrode geometry, are about 15 V and 20 V for the compositionally graded AlN-Al$_{0.73}$Sc$_{0.27}$N and ZnO-Zn$_{0.63}$Mg$_{0.37}$O structures, respectively. These values are significantly smaller than the coercive voltages, 43 V and 40 V, corresponding to the parallel-plate capacitor geometry (as shown in **Fig. 2**). Corresponding "effective" coercive fields (3.8 MV/cm and 5 MV/cm, respectively), are also significantly smaller than those calculated in the parallel-plate capacitor geometry (compare these with the coercive fields of 10.8 MV/cm and 10 MV/cm, respectively). The significant reduction of the coercive voltage appears due to fact that the biased probe acts as a strong external charged defect, and therefore the switching of polarization is polydomain even in the absence of point charge or elastic defects, which are not considered in this work. When the localized field overcomes the coercive field, the domain nucleus emerges under the probe and rapidly intergrows through the structure (see the first top images in **Fig. 4(g)** and **5(g)**), leading to the polarization switching and the hysteresis loop opening. When the bias voltage increases further, the domain walls start to move in the transverse direction. When the bias changes the sign, an oppositely polarized domain starts to grow inside the existing domain (see the bottom images in **Fig. 4(g)** and **5(g)**).

The step-like features at the time sweeps of $\bar{P}_z$ and hysteresis loops of $u_z$ and $\bar{P}_z^{local}$, which are clearly visible in **Figs. 4(c), 4(d)** and **4(f)** (and poorly visible in **Figs. 5(c), 5(d)** and **5(f)**), are associated with the pairwise "annihilation" of domain walls being similar to Barkhausen jumps. As can be seen from **Fig. 4(g)** and **5(g),** the step-like features appear at the loops when the small domain, which grows inside the larger domain, reaches the boundaries of the larger domain. The step-like features of the sweeps and loops are independent of the initial distributions of polarization and strain, but their amount and position depend on the system initial state and geometry. In particular, the steps are absent for the single domain switching in the planar capacitor geometry (see **Fig. 2**).

Let us underline that we observe the pronounced proximity-induced hysteresis loops of the compositionally graded AlN-Al$_{1-x}$Sc$_x$N and ZnO-Zn$_{1-x}$Mg$_x$O structures in the probe-planar electrode geometry, as shown in **Figs. 4(f)** and **5(f)**. Indeed, the polarization of the graded part of the structure, whose chemical composition "x" is close to "otherwise unswitchable" AlN or ZnO, switches simultaneously with the polarization of the part of the structure, which consists of the ferroelectric Al$_{1-x}$Sc$_x$N or Zn$_{1-x}$Mg$_x$O. As can be seen from the images "1" – "10" in **Fig. 4(g)** and **5(g)**, the proximity switching starts in the top AlN and ZnO layers under the PFM probe, where the ferroelectric nanodomains nucleate and intergrow rapidly through the structure. Regardless of what surface (e.g., polar AlN or ferroelectric Al$_{1-x}$Sc$_x$N) contacts the probe, the switching always occurs under the probe, because the electric field is maximal in the region [6].

The proximity-induced local polarization reversal under the probe (as well as the local piezoresponse) in the compositionally graded AlN-Al$_{0.73}$Sc$_{0.27}$N and ZnO-Zn$_{0.63}$Mg$_{0.37}$O structures occur in six stages of nanodomain formation, which are described in detail in Ref. [6] and shown by the images "1" – "10" in **Fig. 4(g)** and **5(g)**. The first is the pre-nucleation stage, when the polarization switching does not occur at low voltages. The second stage is the nucleation stage, when the bias voltage exceeds the coercive value. The third stage is the rapid vertical growth, when the nanodomain elongates and grows vertically, reaches the bottom electrode and forms a cylindrical domain with uncharged domain walls. The fourth stage is the slow lateral growth of the cylindrical domain in the XY-direction with the bias voltage increase. The fifth stage is the back-switching of polarization, which appears when the bias voltage decreases. The nucleus of a new domain appears inside the existing cylindrical domain only when the bias voltage changes its sign and overcomes the critical value, which is determined by the wall pinning effect associated with the finiteness of the numerical grid. The sixth stage is the complete switching of polarization, which occurs when the magnitude of the bias voltage increases further, and the walls of the "nested" domain collapse under the probe.

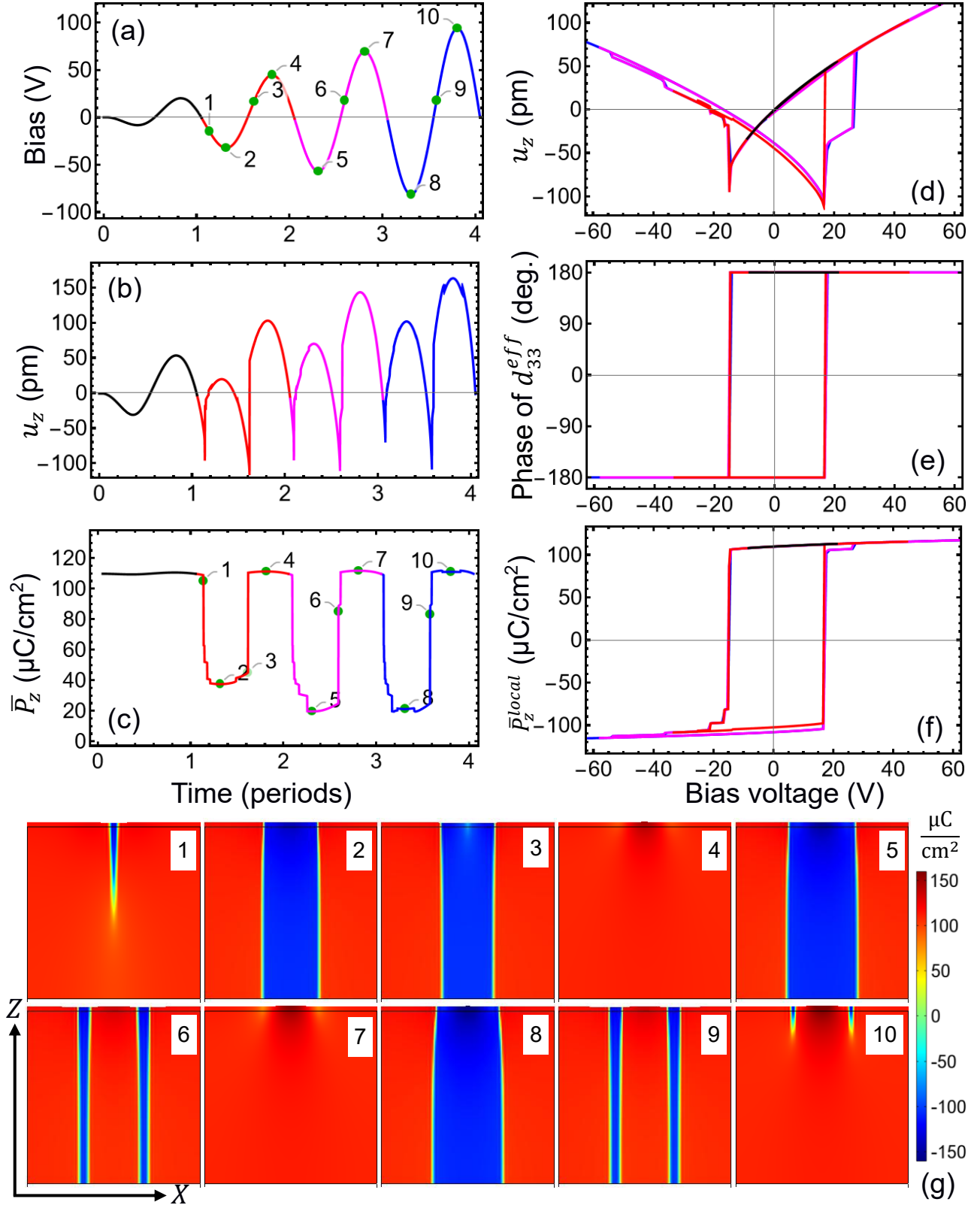

**FIGURE 4**. Time dependences of the bias voltage applied between the PFM probe and the planar electrode **(a)**, the vertical displacement $u_z$ of the surface below the probe tip **(b)** and the average polarization $\bar{P}_z$ **(c)** in the compositionally graded AlN-Al$_{0.73}$Sc$_{0.27}$N structure. Bias voltage dependences of the surface displacement $u_z$ **(d)**, the phase of effective piezoelectric response $d_{33}^{eff}$ **(e)**, and the local polarization $\bar{P}_z^{local}$ **(f)**. **(g)** The distribution of polarization $P_z$ in the XZ-section

of the compositionally graded structure at the moments of time numbered from "1" to "10" shown by the pointers in (a) and (c). The total thickness of the structure $h = 40$ nm, the diffusion length $h_d = 7$ nm, and the tip-surface contact radius $R = 4$ nm. LGD parameters and elastic constants are listed in **Tables S1-S3**.

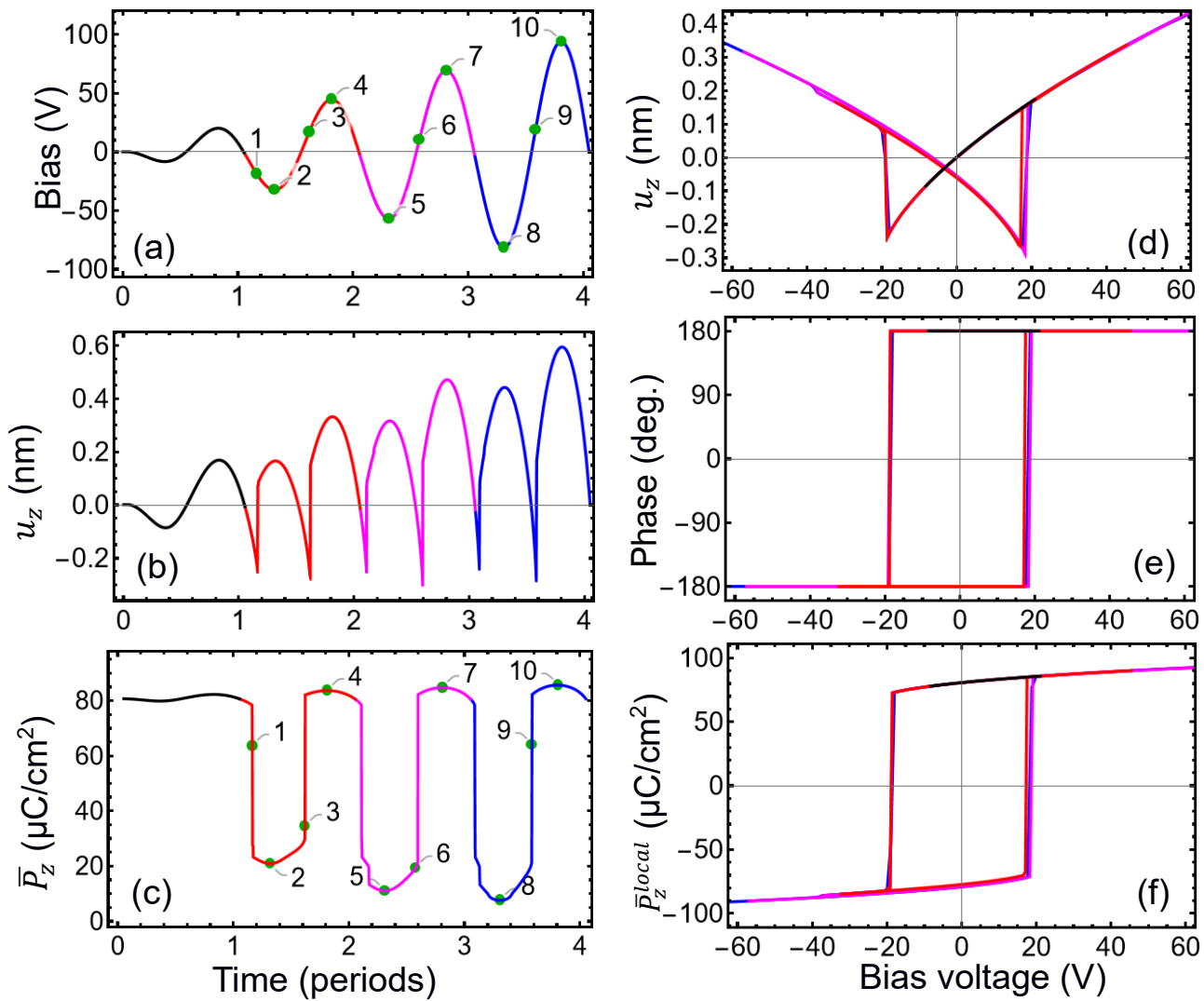

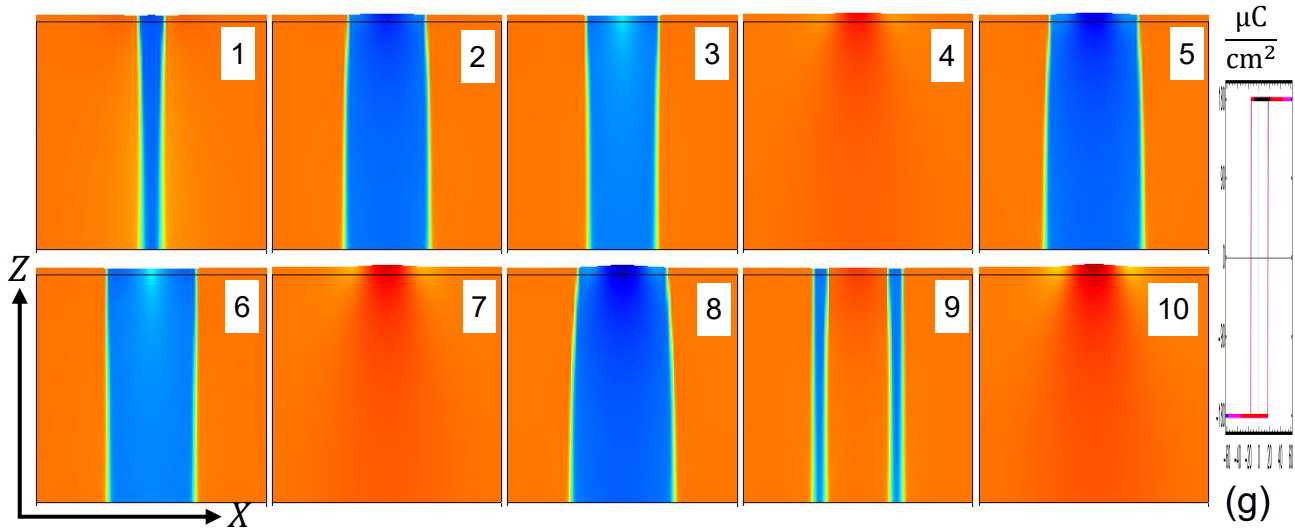

**FIGURE 5**. Time dependences of the bias voltage applied between the PFM probe and the planar electrode **(a)**, the vertical displacement $u_z$ of the surface below the probe tip **(b)** and the average polarization $\bar{P}_z$ **(c)** in the compositionally graded ZnO-Zn$_{0.63}$Mg$_{0.37}$O structure. Bias voltage dependences of the surface displacement $u_z$ **(d)**, the phase of effective piezoelectric response $d_{33}^{eff}$ **(e)**, and the local polarization $\bar{P}_z^{local}$ **(f)**. **(g)** The distribution of polarization $P_z$ in the XZ-section of the compositionally graded structure at the moments of time numbered from "1" to "10" shown by the pointers in (a) and (c). The total thickness of the structure $h = 40$ nm, the diffusion length $h_d = 7$nm, and the tip-surface contact radius $R = 4$ nm. LGD parameters and elastic constants are listed in **Tables S1-S3**.

Notably, consideration of point charged and/or elastic defects can lead to the disappearance of existing step-like features and to a very significant narrowing of the local polarization and piezoresponse hysteresis loop (i.e., to the coercive bias decrease). However, the point defects cannot stabilize charged domain walls with the saw-tooth shape observed recently in the MOCVD Al$_{1-x}$Sc$_x$N films and epitaxial Al$_{1-x}$Sc$_x$N/GaN heterostructures by Wolff et al [27, 28]. The saw-tooth domain fronts "frozen" near the surface, maybe a reflection of the growth non-stoichiometry at high melting temperature, close to the picture observed in the LiNbO$_3$ single-crystals, where the non-stoichiometry can lead to the local variation of the Curie temperature and strong variation of the coercive field [29, 30], at that the non-stoichiometry ratio of Li/Nb may be bigger near the sample surfaces. Also, there are other mechanisms of domain wall pinning that are not considered in the FEM model here due to the small volume of computational cell, such as topological elastic defects (e.g., dislocations) and growth inhomogeneity related with the non-stoichiometry at high growth temperatures. Due to the flexoelectric coupling, the dislocation endings can act as very strong charged defects [31]. Also, we did not consider the band-bending induced by the "bare" depolarization field at the moving charged domain wall, which can lead to the self-screening of the charged domain walls making them metastable.

The physical mechanism of the proximity-induced ferroelectric switching, which is pronounced in the compositionally graded AlN-Al$_{1-x}$Sc$_x$N and ZnO-Zn$_{1-x}$Mg$_x$O structures placed in the probe-planar electrode geometry, is the same as in the parallel-plate capacitor geometry. It is the depolarization electric field determined by the polarization gradient components $\frac{\partial P_z(x,z)}{\partial z}$ and $\frac{\partial P_x(x,z)}{\partial x}$. The field continuously renormalizes the double-well potential of the free energy to lower the steepness of the switching barrier in the "otherwise unswitchable" AlN or ZnO parts of the graded structure under the probe.

Finally, we analyze the polarization switching in the compositionally graded MgO-Zn$_{0.63}$Mg$_{0.37}$O structure placed between the PFM probe and the planar electrode (shown in **Fig. 6**). Since the top side of the structure, whose chemical composition is close to the dielectric MgO at $h < z < h - h_d$, acts as a diffuse dielectric gap, which creates a strong depolarization field, the compositionally graded structure splits into the fine counter-polarized domain stripes far from the probe, which lose their contrast gradually and vanish approaching the top surface $z = h$ (see **Fig. 6(g)**).

The amplitude of the bias voltage increases in time linearly as shown in **Figs. 6(a)**. The time sweep of the vertical surface displacement $u_z$ under the PFM tip and the average polarization $\bar{P}_z$ are shown in **Fig. 6(b)** and **6(c)**, respectively. The time sweep of $\bar{P}_z$ is a slightly anharmonic periodic function corresponding to the polydomain polarization switching and looking like the sweep shown in **Fig. 3(b)**. The sweep of $u_z$ corresponds to a slightly asymmetric and thin double loop of $u_z$ shown in **Figs. 6(d)**. The sweep of $\bar{P}_z$ corresponds the thin and tilted hysteresis loop of the local polarization $\bar{P}_z^{local}$, which has multiple smoothed step-like features and a noticeable coercive voltage ~ 5 V (see **Fig. 6(f)**). Thus, the probe-induced polarization switching corresponds to a noticeable effective coercive field (~0.97 MV/cm), whose existence is confirmed by the rectangular hysteresis loop of the local piezoresponse phase, shown in **Figs. 6(e)**. This result is in contrast to the virtual absence of the coercivity for the polydomain polarization switching in the compositionally graded MgO-Zn$_{0.63}$Mg$_{0.37}$O structure placed in the parallel-plate capacitor geometry (compare **Fig. 6(e)** with **Fig. 3(e)**). The physical origin of the coercivity appearance in the probe-electrode geometry is the formation of larger domains under the probe (compare the width of domain stripes in the images "1", "4", "5" and "8" in **Fig. 6(g)**), as well as the electrostrictive coupling reinforced by the elastic self-clamping in the region of maximal electric field (see the large high-contrast domain under the probe in **Fig. 6(g)**). The domain consolidation and elastic self-clamping are absent in the homogeneous electric field. As expected, the coercive field increases with increase in the ratio $h/h_d$. In contrast, consideration of point charged and/or elastic defects can lead to the very significant narrowing of the local polarization hysteresis loop and to the disappearance of smooth step-like features.

Importantly, the probe-induced domain nucleation and growth start in the MgO-like graded layer under the probe tip (see the images "3" - "10" in **Fig. 6(g)**). Thus, the proximity ferroelectricity in the compositionally graded MgO-Zn$_{0.63}$Mg$_{0.37}$O structure is confirmed by the presence of the ferroelectric domains, which nucleate in the dielectric-like top layer under the probe and intergrow rapidly through the whole structure, as well as by the stability of the counter-polarized domains stripes far from the probe. If the chemical gradient is flipped relative to the tip, the nanodomains nucleate under the probe near the Zn$_{0.63}$Mg$_{0.37}$O surface, because the gradient of the probe field is maximal in the region (see **Fig. S1** in **Supplementary Materials**). Since the depolarization field polarizes the MgO side and depolarizes the Zn$_{0.63}$Mg$_{0.37}$O side of the graded structure, it assists the domain nucleation under the probe and the resulting coercive field is a bit smaller (~0.92 MV/cm) in the structure "tip – compositionally graded Zn$_{0.63}$Mg$_{0.37}$O-MgO – planar electrode" system compared to those in the structure "tip – compositionally graded MgO-Zn$_{0.63}$Mg$_{0.37}$O – planar electrode" (compare **Fig. 6** and **Fig. S1**). Note that the situation for bilayers is opposite, due to the principally different distributions of polarization and electric field in the cases (see **Table S4** and comments to it).

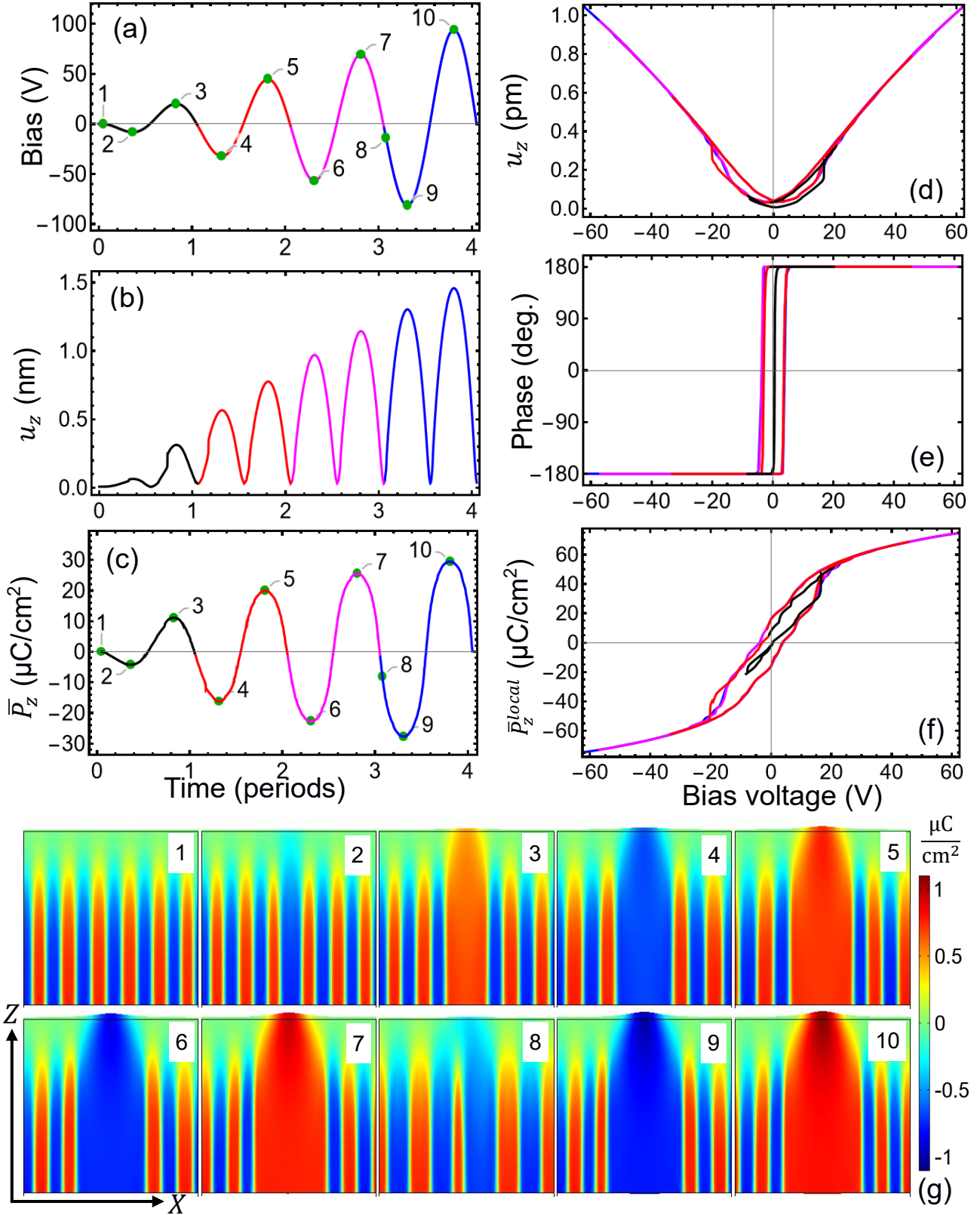

**FIGURE 6**. Time dependences of the bias voltage applied between the PFM probe and the planar electrode **(a)**, the vertical displacement $u_z$ of the surface below the probe tip **(b)** and the average polarization $\bar{P}_z$ **(c)** in the compositionally graded MgO-Zn$_{0.63}$Mg$_{0.37}$O structure. Bias voltage dependences of the surface displacement $u_z$ **(d)**, the phase of the effective piezoelectric response $d_{33}^{eff}$ **(e)**, and the local polarization $\bar{P}_z^{local}$ **(f)**. **(g)** The distribution of polarization $P_z$ in the XZ-section of the compositionally graded structure at the moments of time numbered from "1" to "10" shown by the pointers in (a) and (c). The total thickness of the structure $h = 40$ nm, the diffusion length $h_d = 3$ nm, and the tip-surface contact radius $R = 4$ nm. LGD parameters and elastic constants are listed in **Tables S1-S3**.

The physical mechanism of the proximity ferroelectricity [4], which is pronounced in the compositionally graded MgO-Zn$_{1-x}$Mg$_x$O structure placed in the probe-planar electrode geometry, is the same as in the parallel-plate capacitor geometry. It is the depolarization electric field determined by the polarization gradient components $\frac{\partial P_z(x,z)}{\partial z}$ and $\frac{\partial P_x(x,z)}{\partial x}$. The field continuously renormalizes the parabolic potential of the free energy in the top layer $h < z < h - 3h_d$ towards the appearance of the shallow double-well free energy potential in the "otherwise dielectric" parts of the graded structure under the probe.

## IV. Conclusion

Using LGD thermodynamical approach, we performed the FEM of the polarization, electric and elastic fields evolution in the compositionally graded AlN-Al$_{1-x}$Sc$_x$N, ZnO-Zn$_{1-x}$Mg$_x$O and MgO-Zn$_{1-x}$Mg$_x$O structures sandwiched in the parallel-plate capacitor geometry and in the probe-planar electrode geometry. Coercive fields and spontaneous polarizations of compositionally graded structures, considered in this work, are listed in **Table 1**.

Table 1. Coercive fields and spontaneous polarizations of compositionally graded structures[*]

| Chemical composition of the graded structure | Compositionally graded structure sandwiched between the parallel-plate electrodes | | Compositionally graded structure placed in the probe – planar electrode geometry | |
|---|---|---|---|---|
| | $E_c$, MV/cm | $\bar{P}_S$, µC/cm$^2$ | $E_c$, MV/cm | $\bar{P}_S$, µC/cm$^2$ |
| AlN-Al$_{0.73}$Sc$_{0.27}$N | 10.6 | 109.8 | 4.0 | 109.8 |
| ZnO-Zn$_{0.63}$Mg$_{0.37}$O | 10.0 | 90.1 | 4.8 | 80.7 |
| MgO-Zn$_{0.63}$Mg$_{0.37}$O | 0 | 0 | 1.0 | 20.1 |

[*]The values are calculated for $h = 40$ nm, $R = 4$ nm, $h_d = 7$ nm for the AlN-Al$_{0.73}$Sc$_{0.27}$N and ZnO-Zn$_{0.63}$Mg$_{0.37}$O structures, and $h_d = 3$ nm for the MgO-Zn$_{0.63}$Mg$_{0.37}$O structure. More cases and a comparison with bilayers are listed in **Table S4**.

We revealed that the proximity of the "otherwise unswitchable" polar materials (such as AlN and ZnO) to the switchable ferroelectrics (such as $Al_{1-x}Sc_xN$ and $Zn_{1-x}Mg_xO$) in the compositionally graded structure allows the simultaneous switching of spontaneous polarization in the whole structure by a coercive field significantly lower than the electric breakdown field in the unswitchable polar materials. The physical mechanism of the proximity ferroelectricity [4], predicted here in the compositionally graded AlN-$Al_{1-x}Sc_xN$ and ZnO-$Zn_{1-x}Mg_xO$ structures, is a depolarization electric field determined by the spontaneous polarization gradient in the direction of the chemical composition gradient "x". The field continuously renormalizes the steep double-well potential of the free energy to lower the steepness of the switching barrier in the otherwise unswitchable parts of the graded structure, which chemical composition is close to AlN or ZnO. Hence, the spontaneous polarization of the graded part of the structure, which chemical composition "x" is close to AlN or ZnO, switches simultaneously with the spontaneous polarization of the ferroelectric parts ($Al_{1-x}Sc_xN$ or $Zn_{1-x}Mg_xO$) of the structure.

Also, we predicted that the proximity ferroelectric switching of the compositionally graded AlN-$Al_{1-x}Sc_xN$ and ZnO-$Zn_{1-x}Mg_xO$ structures, placed in the probe-planar electrode geometry, occurs due to the nanodomain formation. The polarization switching starts in the top AlN or ZnO layers placed under the PFM probe, where the ferroelectric nanodomains emerge and intergrow through the whole structure. The coercive voltages and "effective" coercive fields, corresponding to the probe-planar electrode geometry, appear significantly smaller than those calculated in the parallel-plate capacitor geometry, while the hysteresis is highly suppressed. The reduction of the coercive field is due to fact that the biased probe acts as a strong external charged defect. Thus, the switching of the polarization is polydomain even in the absence of point charge or elastic defects, which are not considered in this work. We also can conclude that a chemical composition gradient lowers the effective coercive fields of the compositionally graded AlN-$Al_{1-x}Sc_xN$ and ZnO-$Zn_{1-x}Mg_xO$ structures compared to the coercive fields of the defect-free AlN-$Al_{1-x}Sc_xN$ and ZnO-$Zn_{1-x}Mg_xO$ multilayers with a uniform chemical composition of each layer, where the proximity ferroelectricity was discovered earlier [4, 5, 6].

We observed the proximity ferroelectric switching in the compositionally graded MgO-$Zn_{1-x}Mg_xO$ structure, which physical mechanism is the depolarization electric field also. The field continuously renormalizes the parabolic free energy potential in the graded MgO-like layer towards the appearance of the shallow double-well free energy potential. In this case the depolarization field, reinforced by the gradient of chemical composition "x", induces the formation of thermodynamically stable of low-contrast or high-contrast domain stripes in the different parts of the compositionally graded structure. The compositionally graded MgO-$Zn_{1-x}Mg_xO$ structure sandwiched in a parallel-plate capacitor has a much smaller effective coercive field and AFE-like polarization hysteresis, indicating on low power consumption with complete switching. Such type of poly-domain polarization switching leads to the virtual absence of coercivity, at the same time allowing to avoid energy dissipation, while being able to switch.

A hysteresis loop with a pronounced coercive voltage appears in the compositionally graded MgO-Zn$_{0.63}$Mg$_{0.37}$O structure placed in the probe-electrode geometry, while the loop is absent in the parallel-plate capacitor geometry. Thus, the proximity ferroelectricity in the compositionally graded MgO-Zn$_{0.63}$Mg$_{0.37}$O structure is confirmed by the presence of the ferroelectric domains, which nucleate in the dielectric-like top layer under the probe and intergrow through the whole structure, as well as by the stability of the counter-polarized domains stripes far from the probe.

Using proximity ferroelectricity, the tip-induced control of ferroelectric domains in otherwise nonswitchable compositionally graded structures can provide nanoscale domain engineering for memory, actuation, sensing and optical applications.

**Acknowledgements.** The work is supported by the DOE Software Project on "Computational Mesoscale Science and Open Software for Quantum Materials", under Award Number DE-SC0020145 as part of the Computational Materials Sciences Program of US Department of Energy, Office of Science, Basic Energy Sciences. The work of A.N.M. is funded by the National Research Foundation of Ukraine (project "Silicon-compatible ferroelectric nanocomposites for electronics and sensors", grant N 2023.03/0127). The work of E.A.E. is funded by the National Research Foundation of Ukraine (project "Manyfold-degenerated metastable states of spontaneous polarization in nanoferroics: theory, experiment and perspectives for digital nanoelectronics", grant N 2023.03/0132). Obtained results were visualized in Mathematica 14.0 [32].

## Supplementary Materials to the Manuscript

LGD parameters of Al$_{1-x}$Sc$_x$N and Zn$_{1-x}$Mg$_x$O, used in the FEM, are listed in **Table SI.** The LGD parameters of Al$_{1-x}$Sc$_x$N were determined from the experimentally measured spontaneous polarization [33, 34] and linear dielectric permittivity [35] as described in Refs. [36] and [5]. The background permittivity $\varepsilon_b$ is estimated as the square of refractive index according to Ref. [37]. The LGD parameters of Zn$_{1-x}$Mg$_x$O were determined from the experimental results of Ferri et al. [38], who measured both the quasi-static dielectric constant at low field and the ferroelectric hysteresis loops in Zn$_{1-x}$Mg$_x$O. The spontaneous polarization of ZnO can be estimated from the first principles calculations [39] being close to 0.9 C/m$^2$, and its dielectric permittivity value was taken from the experimental results of Hofmeister et al. [40]. The dielectric parameters of the chemically pure cubic MgO were estimated from its dielectric permittivity $\varepsilon_{stat} \approx$9.83 and $\varepsilon_b \approx$ 3.02 [41]. Also, we assume equal gradient tensor coefficients in longitudinal and transverse directions.

Table S1. LGD model parameters of $Al_{1-x}Sc_xN$ and $Zn_{1-x}Mg_xO$ *

| compound | $\alpha_i$, m/F | $\beta_i$, m$^5$/(F C$^2$) | $\gamma_i$, m$^7$/(F C$^4$) | $E_C^{bulk}$, MV/cm** | $E_C^{film}$, MV/cm*** | $\varepsilon_b^{(i)}$ |
|---|---|---|---|---|---|---|
| AlN layer | $-2.164 \cdot 10^9$ | $-3.155 \cdot 10^9$ | $2.788 \cdot 10^9$ | 26.1 | 24.8 | 2.1 |
| $Al_{0.73}Sc_{0.27}N$ | $-2.644 \cdot 10^8$ | $-3.155 \cdot 10^9$ | $2.788 \cdot 10^9$ | 9.3 | 7.4 | 2.2 |
| ZnO | $-7.152 \cdot 10^9$ | $8.829 \cdot 10^9$ | 0 | 24.8 | 22.5 | 3.40 |
| $Zn_{0.63}Mg_{0.37}O$ | $-3.870 \cdot 10^8$ | $-2.914 \cdot 10^9$ | $2.8 \cdot 10^9$ | 8.9 | 0.65 | 2.42 |
| MgO | $1.659 \cdot 10^{10}$ | 0 | 0 | 0 | 0 | 3.02 |

* The polarization gradient coefficient is taken the same for all materials, $g_{z,\perp}^{(i)} = 1.0 \cdot 10^{-10}$ m$^3$/F

** The thermodynamic coercive field of a stress-free bulk material

*** The thermodynamic coercive field of a thin epitaxial film clamped to a rigid substrate. Dislocations or/and any other sinks of elastic strains are absent.

Electrostriction coefficients $Q_{ij}$ and elastic stiffness $c_{ij}$ tensors components of ferroelectric, polar and/or dielectric layers, which are regarded the same, are collected from Refs. [24, 42]. They are listed in **Table S2** and **S3.** We use a uniform rectangular mesh with the size 2 nm, and the linear size of the cubic computation cell is 40 nm.

Table S2. Elastic parameters of AlN and $Al_{1-x}Sc_xN$

| parameters | $Q_{ij}$, m$^4$/C$^2$ | Ref. | $c_{ij}$, GPa | Ref. |
|---|---|---|---|---|
| AlN | $Q_{13} = -0.0087$, $Q_{33} = 0.0203$ | [24] | $c_{11}$=396, $c_{12}$=137, $c_{13} = 108$, $c_{33} = 373$, $c_{44} = 116$, $c_{66} = 130$ | [43] |
| $Al_{0.73}Sc_{0.27}N$ | $Q_{13} = -0.0152$, $Q_{33} = 0.0406$ | [24] | $c_{11}$=319, $c_{12}$=151, $c_{13} = 127$, $c_{33} = 249$, $c_{44} = 101, c_{66} = 84$ | [43] |

Table S3. Elastic parameters of ZnO, MgO and $Zn_{1-x}Mg_xO$

| material | parameters | units | values | Ref. |
|---|---|---|---|---|
| ZnO | stiffness | GPa | $c_{11}$=209, $c_{12}$=120, $c_{13}$=104, $c_{33}$=218, $c_{44} = 44.1$ | [44] |
| ZnO | piezoelectric coefficients | pm/V | $d_{31} = -5.43, d_{33} = 11.67, d_{15} = -11.34$ | [45] |
| ZnO | electrostriction | m$^4$/C$^2$ | $Q_{13} = -0.043, Q_{33}$=0.093, $Q_{55}$=-0.269 | * |
| MgO | stiffness | GPa | $c_{11}$=294, $c_{12}$=93, $c_{44} = 155$ | [46] |
| MgO | electrostriction | m$^4$/C$^2$ | $Q_{12} = Q_{13} = -0.09, Q_{11} = Q_{33} = 0.34, Q_{11} + 2Q_{12} = 0.16$ | [47], [48] |
| $Zn_{0.63}Mg_{0.37}O$ | electrostriction | m$^4$/C$^2$ | $Q_{13} = -0.06, Q_{33}$=0.18, | ** |
| $Zn_{0.63}Mg_{0.37}O$ | stiffness | GPa | $c_{11}$=240, $c_{12}$=110, $c_{13}$=100, $c_{33}$=246, $c_{44} = 85$ | *** |

*The relationship between the piezoelectric and electrostriction coefficients have the following form (see e.g. Ref. [49]): $Q_{13} = \frac{d_{31}}{2\varepsilon_0(\varepsilon_{33}-\varepsilon_b)P_s}$, $Q_{33} = \frac{d_{33}}{2\varepsilon_0(\varepsilon_{33}-\varepsilon_b)P_s}$, $Q_{55} = \frac{d_{15}}{\varepsilon_0(\varepsilon_{11}-\varepsilon_b)P_s}$.

**Recalculated using the linear mixing rule of ZnO and MgO electrostriction constants

***Recalculated using the linear mixing rule of ZnO and MgO elastic constants

Coercive fields and spontaneous polarizations of the compositionally graded structures and bilayers, listed in **Table S4**, are calculated for the total thickness $h = 40$ nm, tip-surface contact radius $R = 4$ nm, diffusion length $h_d = 7$ nm (for the compositionally graded AlN-Al$_{0.73}$Sc$_{0.27}$N and ZnO-Zn$_{0.63}$Mg$_{0.37}$O structures) or $h_d = 3$ nm (for the compositionally graded MgO-Zn$_{0.63}$Mg$_{0.37}$O and Zn$_{0.63}$Mg$_{0.37}$O-MgO structures), $h_2 = 20$ nm (for the AlN-Al$_{0.73}$Sc$_{0.27}$N and ZnO-Zn$_{0.63}$Mg$_{0.37}$O bilayers) and $h_2 = 9$ nm (for the MgO-Zn$_{0.63}$Mg$_{0.37}$O and Zn$_{0.63}$Mg$_{0.37}$O-MgO bilayers). Since $\exp(-h/h_d) \approx 0.05$ for $h = 3h_d$, the estimate $3h_d \cong h_2$ seems valid.

**Table S4.** Coercive fields and spontaneous polarizations of the bilayers and compositionally graded structures

| Composition | Bilayer sandwiched between the parallel-plate electrodes | | Compositionally graded structure sandwiched between the parallel-plate electrodes | | Bilayer placed in the probe – planar electrode geometry | | Compositionally graded structure placed in the probe – planar electrode geometry | |
|---|---|---|---|---|---|---|---|---|
| | $E_c$, MV/cm | $P_s$, μC/cm² | $E_c$, MV/cm | $P_s$, μC/cm² | $E_c$, MV/cm | $P_s$, μC/cm² | $E_c$, MV/cm | $P_s$, μC/cm² |
| AlN-Al$_{0.73}$Sc$_{0.27}$N | 17.7 | 119.9 | 10.6 | 109.8 | 5.8 | 117.0 | 4.0 | 109.8 |
| ZnO-Zn$_{0.63}$Mg$_{0.37}$O | 14.7 | 95.9 | 10.0 | 90.1 | 5.9 | 81.6 | 4.8 | 80.7 |
| MgO-Zn$_{0.63}$Mg$_{0.37}$O | 0 | 0 | 0 | 0 | 1.53* | 0 | 0.97 | 20.1 |
| Zn$_{0.63}$Mg$_{0.37}$O-MgO | 0 | 0 | 0 | 0 | 0.47 | 6.8 | 0.92 | 27.0 |

*The field is in fact the first critical field of a double loop opening. The single loop and the spontaneous polarization are absent in this case. The absence of a single loop in the MgO-Zn$_{0.63}$Mg$_{0.37}$O bilayer and the appearance of a relatively thin relaxor-like single loop in the Zn$_{0.63}$Mg$_{0.37}$O-MgO bilayer may be related to the strong weakening of the tip field by the 9-nm thick MgO layer. At the same time the field initiates easily the domain nucleation at the surface of Zn$_{0.63}$Mg$_{0.37}$O layer, when the layer is placed under the probe.

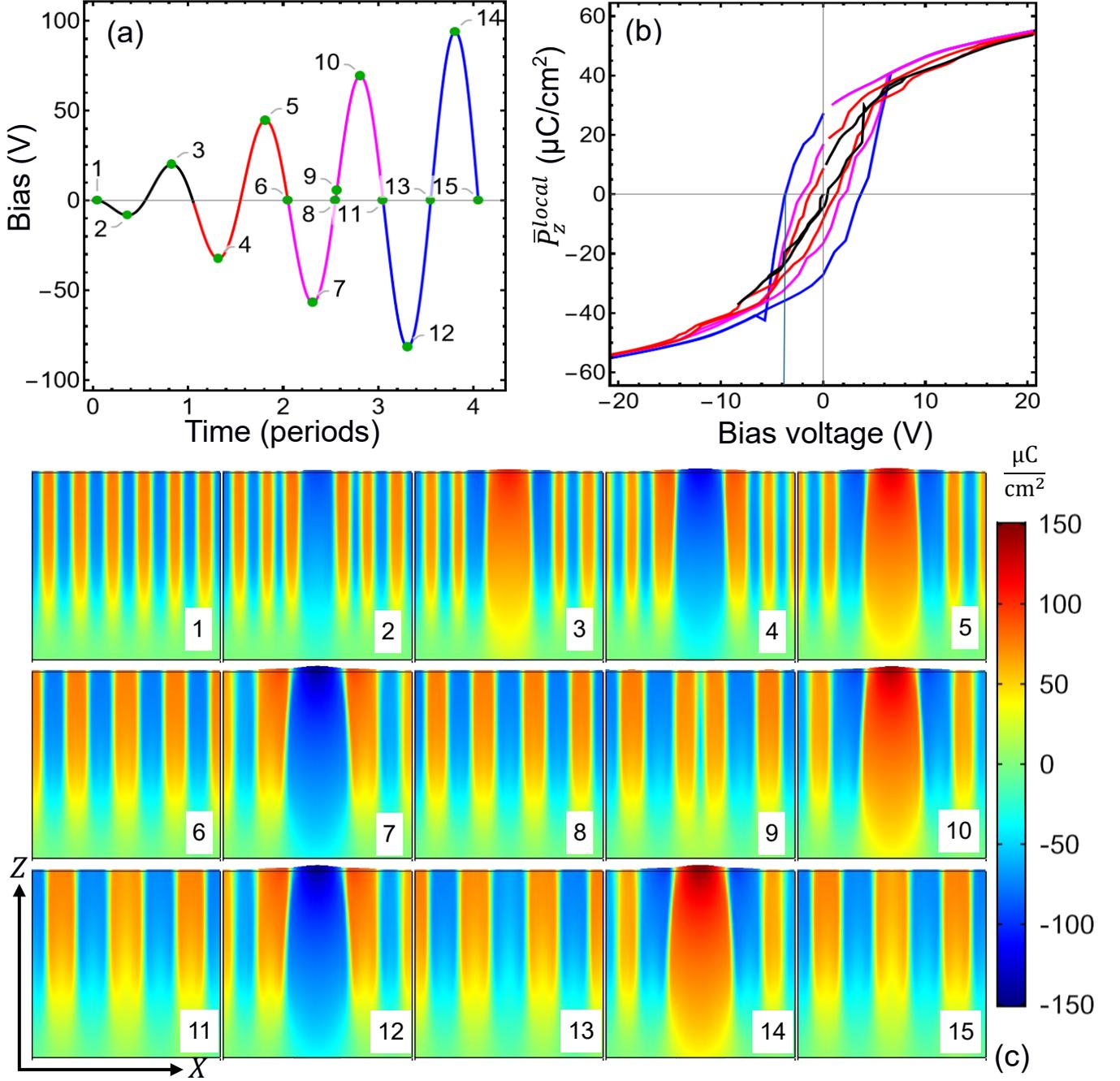

**FIGURE S1.** **(a)** Time dependences of the bias voltage applied between the PFM probe and the planar electrode in the compositionally graded $Zn_{0.63}Mg_{0.37}O$ - MgO structure. **(b)** Bias voltage dependences of the local polarization $\bar{P}_z^{local}$. **(c)** The distribution of polarization $P_z$ in the XZ-section of the compositionally graded structure at the moments of time numbered from "1" to "15" shown by the pointers in **(a)**. The total thickness of the structure $h = 40$ nm, the diffusion length $h_d = 3$ nm, and the tip-surface contact radius $R = 4$ nm. LGD parameters and elastic constants are listed in **Tables S1-S3**.